\documentclass[11pt]{article}
\usepackage[utf8]{inputenc}
\pdfoutput=1
\usepackage{amssymb,amsmath,mathrsfs,enumerate}
\usepackage{graphicx,rotate,multicol}
\usepackage{float}
\usepackage{tocloft}
\usepackage{subfig}
\usepackage[margin=10pt,labelfont=bf]{caption}
\usepackage{cite}
\usepackage{soul}
\usepackage[colorlinks=true,
            linkcolor=blue,
            urlcolor=blue,
            citecolor=teal]{hyperref}
\usepackage{color}
\definecolor{ourcolor}{rgb}{0.7, 0.25, 0.05}

\usepackage{tikz,braket}
\long\def\rpl#1!!#2!!{\textcolor{red}{#1} \textcolor{blue}{#2}}

\let\tilde=\widetilde

\let\bar=\overline

\def \order(#1){{\mathcal O} \left(#1 \right)}

\evensidemargin=\oddsidemargin
\allowdisplaybreaks

\title{\color{black}{\bf Neutrino Condensate Dark Energy from TeV Scale Extra Dimensions}}

\author{\bf Ujjal Kumar Dey,$^{a,}$\footnote{ujjal@iiserbpr.ac.in} 
\hspace{4pt} ~and~  Utpal Sarkar$^{b}$\footnote{utpal.sarkar@iiserkol.ac.in} \\[10pt]
\small\em $^a$Department of Physical Sciences, Indian Institute of Science Education and Research Berhampur,\\ 
\small\em Transit Campus, Government ITI, Berhampur 760010, Odisha, India \\
\small\em $^b$Department of Physical Sciences, Indian Institute of Science Education and Research Kolkata,\\
\small\em Mohanpur 741246, India
}

\date{}

\begin{document}

\maketitle

\begin{abstract}
We propose a new scenario of incorporating neutrino masses in models of TeV scale gravity and large extra dimensions, which can explain the dark energy problem through formation of neutrino condensates. The smallness of the neutrino masses and the scale of dark energy is protected to be small by the lepton number symmetry, which is broken in a distant brane, such that all lepton number violating effects are suppressed by the distance factor in the extra dimensions. The interactions of any bulk scalar particles are also suppressed by the distance factor. A lepton number violating soft-term is also induced from the distant brane, providing us a light pseudo-Nambu-Goldstone boson, which becomes the mediator of the attractive force that forms the condensates. 
\end{abstract}

%~~~~Keywords: Beyond Standard Model; Rare decay; FCNC Interaction

%\newpage
%
%\hrule \hrule
%\tableofcontents
%\vskip 10pt
%\hrule \hrule 

\vskip 30pt

The cosmic microwave anisotropy analysis and related astrophysical observations established the energy budget of our universe, and showed that about 70\% of the total energy consists of dark energy~\cite{Planck:2015fie}. This dark energy posed a major challenge to models of particle physics, because unlike ordinary matter (visible or dark matter), the dark energy provides negative pressure forcing the universe to expand in an accelerated manner. Moreover, this accelerating phase has started only in the present epoch, which also needs an explanation. Several models were proposed to understand
the dark energy problem and their phenomenological consequences were also studied ({see e.g.,~\cite{Wetterich:1987fm, Ratra:1987rm, Copeland:2006wr, Bamba:2012cp, Yoo:2012ug, doi:10.1142/8906}}). In a class of model, neutrinos play a crucial role
\cite{Gu:2003er, Fardon:2003eh, Peccei:2004sz, Li:2004tq, Barger:2005mn, Brookfield:2005bz, Brookfield:2005td, Ringwald:2006ks, Barbieri:2005gj, Takahashi:2005kw, Fardon:2005wc, Ichiki:2008st, Ma:2006mr, Bhatt:2008hr, Franca:2009xp, Ghalsasi:2014mja, Geng:2015haa}.
Another possibility of explaining the dark energy is to consider the neutrino
(or some other neutral fermions) to condensate
\cite{Antusch:2002xh, Kapusta:2004gi, Yajnik:2005pr, Bhatt:2009wb, Yajnik:2011mh, Dvali:2016uhn,  Inagaki:2016vkf, Addazi:2016oob, Gu:2006dc, Chodos:2020dgi, Ma:2000xh, Azam:2010kw, Nollett:2014lwa, Davidson:2009ha}.
In one of the recent proposals, it was pointed out that the neutrinos can form condensates in the presence of a light scalar field providing the attractive interaction of the neutrinos, which can explain the accelerating universe. Moreover, since the scale of neutrino masses is similar to our present background microwave radiation temperature (about a meV), this scenario explains why our universe enters into the accelerating phase only recently. 
The starting point for the neutrino condensate models is an interaction between the neutrinos provided by a scalar field, which gives rise to the required attractive force between the neutrinos. This interaction could be of the form:
\begin{eqnarray}
 {\cal L}_{\nu S} &=& f_{S \alpha \beta} ~
 \nu_{\alpha R} \nu_{\beta R} S + M_{S \alpha \beta} ~
 \nu_{\alpha R} \nu_{\beta R}, \nonumber \\
 {\cal L}_{\nu D} &=& f_{D i \alpha } ~
 \bar{\nu}_{i L} \nu_{\alpha R} \eta + M_{D i \alpha} ~
 \bar{\nu}_{i L} \nu_{\alpha R}, \nonumber \\
 {\cal L}_{\nu T} &=& f_{M i j } ~
 \nu_{i L} \nu_{j L } \xi + M_{M i j} ~
 \nu_{i L} \nu_{j L}.  
\end{eqnarray}
The first one is the singlet interaction (${\cal L}_{\nu S}$), where $S$ is an $SU(2)_L$ singlet scalar field that mediates the interaction between two right-handed or sterile neutrinos (we denote both right-handed neutrinos and sterile neutrinos as $\nu_R$). The vacuum expectation values (vev) of any singlet scalar field can give rise to the Majorana mass term for the $\nu_R$. The second one gives the doublet interaction (${\cal L}_{\nu D}$), where $\eta$ is an $SU(2)_L$ doublet scalar field that mediates the interaction. If $\eta$ gets an usual Dirac see-saw small vev, it can provide the Dirac neutrino mass to the neutrinos coupling the left-handed fields with the right-handed fields. The third one is the triplet interaction (${\cal L}_{\nu T}$) mediated by an $SU(2)_L$ triplet scalar field ($\xi$), which can provide a Majorana mass to the left-handed neutrinos. 
The original proposal for the neutrino condensate dark energy worked with a singlet interaction and a singlet scalar field ($S$) could provide the attractive interaction for the right-handed and sterile neutrinos. So, the connection between the observed left-handed neutrino masses and the dark energy was indirect. This problem is remedied in a recent work, where the attractive force between the left-handed and right-handed neutrinos originates from the doublet interaction, and hence, the neutrinos could get only Dirac mass. The third possibility of neutrino condensates mediated by the triplet Higgs scalar $\xi$ will truly connect the neutrino mass generation with the dark energy and the resultant neutrinos will be Majorana particles. So, these models have the prospect of incorporating leptogenesis in an extended version. 
In this article we shall study the dark energy through 
triplet interaction, which has some similarity to
the one with doublet interaction. The main ingredient for
forming neutrino condensate is a light scalar field which
gives the attractive force~\cite{Kapusta:2004gi, Chodos:2020dgi}. On the other hand, the 
interactions of any light
scalar field with the standard model Higgs doublet 
would tend to make it heavy, unless protected by any
symmetry. This could be achieved when the model has
a Lepton number $U(1)_L$ symmetry and a scalar singlet field becomes
a pseudo-Nambu-Goldstone-Boson (pNGB). Thus a singlet scalar
field becomes a natural choice for the mediator of 
neutrino condensate. In the present case of left-handed
Majorana neutrino condensate
model with the triplet interaction, both the singlet 
($S$) and the triplet scalar ($\xi$) breaks the $U(1)_X$ 
symmetry and as a result, the pNGB contains both
the singlet and triplet scalar components. The mixing
of the neutral imaginary component of the triplet 
with the singlet pNGB then provides the attractive
force between the two left-handed Majorana neutrinos and
form the condensate.

The next challenge in explaining the dark energy is
to achieve the scale of the neutrino condensate,
which should be related to the scale of neutrino masses
and protected by some symmetry.
On the other hand, there should not be unwanted light
scalars which could affect the cosmic evolution or
invisible Higgs decay. In the
present model all these could be ascertained by embedding 
the present scenario in models with low scale gravity
and extra dimensions.

%\section{Neutrino Condensate Model in Large Extra Dimension}

The basic model is similar to the usual triplet Higgs model
of neutrino masses from extra dimension and distant
symmetry breaking mechanism~\cite{Ma:2000xh}. 
We modify the model suitably to explain the 
neutrino condensate dark energy. The main ingredient
is a new lepton number violating soft term originating from
an interaction in the distant brane, which naturally
gives a tiny mass to the lepton number violating 
Nambu-Godstone Boson, making it a pseudo-Nambu-Goldstone
Boson (pNGB). This pNGB can then provide the interactions between
the neurinos that gives the attractive force for 
forming the neutrino condensates. 

Another crucial input in the model is the
interaction strength of the pNGB. The interactions of
the pNGB with the leptons is suppressed by the $vev$s
of the singlet and triplet Goldstone modes, but since the
singlet scalar propagates mostly in the bulk, the mixing of
the pNGB with the singlet scalar is also suppressed, and
hence, the mixing of the pNGB with the leptons are 
proportionately enhanced, allowing the model to predict
the correct amount of dark energy. 

We consider a minimal extension of the standard model by
extending only the scalar sector with a triplet Higgs
scalar $\xi$ and a singlet Higgs scalar $\chi$:
\begin{eqnarray}
\xi =\left(\begin{array}{cc} \xi^+/\sqrt{2} & \xi^{++}\\ 
-\xi^0 & -\xi^+/\sqrt{2} \end{array}\right)\sim (3,1)\,,
~~~~~~~~~\chi=\chi^0\sim (1,0)\,.
\end{eqnarray}
We assume that lepton number is conserved in our world
and both these fields carry lepton number $-2$, so that
the interactions of these fields (in addition to the usual
self-energy and quartic interaction terms) are given
by
\begin{equation}
{\cal L}_Y= f_{ij} L_i^T C^{-1}\, i\tau_2\, \xi\, L_j + 
\lambda \Phi^\dagger \xi \tilde\Phi \chi^\dagger (x, y=0) +
h.c.\,.
\end{equation}
$\Phi$ is the standard model Higgs doublet and 
the Higgs triplet $\xi$
is written in a $2\times 2$ matrix representation, 
transforming under $SU(2)_L$
as $\xi \to U\xi U^\dagger.$ 

The singlet field $\chi$ plays a crucial role in
this model. We assume that our world is embedded
in a higher dimensional space. In the simplest scenario,
we consider a $(4+n)-$dimensional model with coordinates
$x_i, ~i=0,1,2,3$ and $y_k,~k=1,\ldots n$, 
and our 3-brane ${\cal P}$ ({\it i.e.,} our
4-dimensional world) is localized at $y=0$ in the
fifth dimension. Only the singlet scalar
$\chi$ can propagate along the bulk ($y-$directions, so
its position is given by both $x$ and $y$), whereas all
the usual standard model particles reside in
our 4-dimensional world at $y=0$.

There exists another 3-brane ${\cal P}'$
at the other end of the bulk at $y=y_*$, where lepton
number is violated spontaneously by the vacuum expectation value ($vev$) of
a singlet scalar field $\eta$ carrying lepton number $(-2)$. 
The messenger field $\chi$ interacts with $\eta$ through the 
interaction
\begin{equation}
{\cal S}_{\rm other} = \int_{{\cal P}'} d^4 x' ~\mu^2~ \eta (x') \chi 
(x', y = y_*), 
\end{equation}
where $\mu$ is a mass parameter. 
The spontaneous violation of lepton number at ${\cal P}'$ 
will be carried to our brane by the messenger field $\chi$
and the effect will be suppressed by the distance
between the branes ${\cal P}$ and ${\cal P}'$, which may
be estimated as 
\begin{equation}
|y_*| = r = M_P^{2/n}~/ ~ M_*^{1+2/n} \,,
\end{equation}
where $M_P$ is the reduced Planck scale $2.4 \times 10^{18}$~GeV
and $M_*$ is some fundamental scale.

Assuming $\langle \eta \rangle$ to be a point source in
the brane ${\cal P}'$, lepton number 
violation in our brane ${\cal P}$ 
may be estimated ~\cite{Ma:2000xh} by the ``shined" value of $\langle \chi \rangle$,
\begin{equation}
\langle \chi (x, y = 0) \rangle = \Delta_n (r) \langle \eta (x, y = y_*) 
\rangle,
\end{equation}
where $\Delta_n(r)$ is 
the Yukawa potential in $n$ transverse dimensions, i.e.
\begin{equation}
\Delta_n (r)= {1 \over (2 \pi )^{n \over 2}
	M_*^{n- 2}} ~\left( {m_\chi \over r} \right)^{n-2 \over 2}
~K_{n - 2 \over 2} \left( m_\chi r \right),
\end{equation}
$K$ being the modified Bessel function.  If the mass of $\chi$ 
is large ($m_\chi r \gg 1$), the suppression is exponential, 
\begin{eqnarray} 
\langle \chi \rangle \approx \displaystyle{ 
	m_\chi^{n - 3 \over 2} \over 2 (2 \pi)^{n - 1 
		\over 2} M_*^{ n - 2} } \displaystyle{e^{- 
		m_\chi r} \over r^{n-1 \over 2} }.
\end{eqnarray}
Whereas, if the mass of $\chi$ is small and $n>2$, 
the ``shined" value of $\chi$ is 
\begin{equation}
\langle \chi \rangle \approx { \Gamma ( {n -2 \over 2} ) \over
	4 \pi^{n \over 2} }{M_* \over (M_* r)^{n-2} }.
\label{chivev}
\end{equation}
Thus the lepton number
violating effects become naturally suppressed in
our brane and the smallness of the neutrino masses
has a natural explanation. 

There is a
massless Nambu-Goldstone Boson in the model in our brane, 
the Majoron,
arising from the spontaneous breaking of the lepton number
symmetry $U(1)_L$ by the ``shined" value of the
singlet scalar $\chi$. At this stage we deviate from the
earlier model of the triplet Higgs scalar and introduce a
new interaction in the distant brane, which will induce
a soft lepton number violating term in our brane. This will, in turn,
provide a small mass to the Majoron, 
and make it a pseudo-Nambu-Goldstone
boson (pNGB). The lepton number symmetry will protect the
smallness of the pNGB mass making it a natural choice
for the scalar field that gives rise to the attractive
force between the neutrinos forming condensates. 

The soft explicit lepton number violating interaction in our brane
originates from 
interactions of $\chi$ with the field $\eta$ in
the extra dimensions making the pNGB light with mass of order $10^{-3}$ eV. 
The relevant term that allows the explicit
lepton number breaking in our brane and makes the Majoron
a pNGB originates from a lepton number conserving
term in the other brane at $y=y_*$:
\begin{equation}
S_{\rm exp} = \int_{{\cal P}'}~ d^4 x' ~ \lambda_{\rm exp} \text{Tr}
[\eta (x') \eta (x') ] ~
\text{Tr}[\chi (x',y=y_*)][\chi (x', y=y_*)]  
\end{equation}
The $vev$ of the field $\eta$ breaking lepton number in the
distant brane will then introduce an
explicit lepton number violating soft term in the
scalar potential in our brane, 
\begin{equation}
V_{\rm exp} = m_s^2 ~\chi^2  = 
[ \Delta_n(r) \langle \eta \rangle (x, y=y_*)]^2
~\chi^2 
\end{equation}
The pNGB will now have a tiny mass $m_s$
and the lepton number symmetry will protect this mass from
any large corrections. This will eliminate the naturalness problem 
of the dark energy and also explain why the universe starts
accelerating in the present epoch. 

We shall now study the mass spectrum of the
scalars. Since this lepton number violating
term $V_{\rm exp}$ will not affect our present discussions on the mass
spectrum, except giving a small mass $m_s$ to the Majoron
making it a pNGB, we shall include the contribution of 
this term at a later stage.
We assume that the $vev$ of the bulk field $\chi$ is
determined by this ``shined" value ($\langle \chi \rangle 
= z / \sqrt{2} $) and we express this field as
\begin{equation}
\chi = {1 \over \sqrt 2} (\rho + z) e^{i\varphi}\,,
\end{equation}
and the dynamics in our brane will not affect this field in any
way and $z$ appears as a boundary condition. The Higgs potential
of this model has been studied extensively~\cite{Ma:2000xh}, and we shall 
use the results that are relevant for our discussions. 

The new ingredient in our model comes from the understanding
of the propagator of the bulk scalar. Similar to the graviton,
the scalar field $\chi$ propagates mostly in the bulk and its
interactions with the field in our brane are suppressed by
the volume of the bulk $V$. The value of the $vev$ $z$ has
been determined including the effect of the bulk volume $V$,
so there is no change. The physical field $\rho$ is the
heavy particle, whose
interactions will be scaled by the volume of the bulk $V$, but
it will not affect our present analysis, so we shall not make
the corresponding changes. 

The pseudoscalar normalized physical field $z\varphi$ will
also be scaled by the bulk volume $V$, which plays a crucial
role in this model. We shall thus express the above equation
as
\begin{equation}
\chi = {1 \over \sqrt 2} (\rho + z) e^{i\varphi/V}\,,
\end{equation}

Various parameters are determined from the
phenomenological considerations. The $vev$ of $\chi$ will
break the lepton number spontaneously in our brane leading
to a Majoron ($ z \varphi$). The lepton number transformations 
of the various fields in our brane are
\begin{equation}
\rho\to \rho\,,~~~~ \varphi\to\varphi - 2x\,,~~~~
\nu\to e^{ix} \nu\,, ~~~~ \xi\to e^{-2ix} \xi\,.
\end{equation}
We can now write down the lepton number conserving 
Higgs potential in our brane as
\begin{eqnarray}
V &=& m_0^2 \,\Phi^\dagger \Phi + m_\xi^2\, \rm{Tr}[\xi^\dagger \xi] + 
{1 \over 2} \lambda_1 (\Phi^\dagger \Phi)^2 + 
{1 \over 2} \lambda_2 \text{Tr}[\xi^\dagger \xi]^2 + 
\lambda_3 (\Phi^\dagger \Phi)\text{Tr}[\xi^\dagger \xi] \nonumber \\ 
&& + \lambda_4 \text{Tr}[\xi^\dagger \xi^\dagger]\text{Tr}[\xi \xi]
+ \lambda_5 \Phi^\dagger \xi^\dagger \xi \Phi +
\left( {\lambda_0 z e^{-i\varphi} \over \sqrt 2} 
\Phi^\dagger \xi \tilde \Phi + h.c. \right) \,,
\label{V}
\end{eqnarray}
where $m_0^2 < 0$ and $m_\xi^2 > 0$. But the effective trilinear
interaction in our brane (the last term ) will induce
a $vev$ to the the triplet Higgs scalar $\langle \xi^0 \rangle
= u/\sqrt{2}$, after the usual Higgs doublet acquires a 
$vev$ $\langle \phi^0 \rangle = v/\sqrt{2}$. Expressing 
\begin{equation}
\phi^0 = {1 \over \sqrt 2} (H + v) e^{i\theta}, ~~~~ 
\xi^0 = {1 \over \sqrt 2} (\zeta + u) e^{i\eta},
\end{equation}
we can minimize the potential to find out the $vev$s:
$$ v^2 \simeq -2m_0^2/\lambda_1 \,, ~~~{\rm and}~~~ u \simeq \lambda_0 
z v^2/2m_\xi^2 \,. $$
The $vev$ ($u$) of the triplet scalar thus becomes very small 
naturally and gives the required neutrino masses
\begin{equation}
M^\nu_{ij} = 2 f_{ij} \langle \xi \rangle = 
\frac{1}{\sqrt{2}} f_{ij} \lambda_0 z 
\frac{v^2}{m^2_{\xi}} \,.
\label{eq:numass}
\end{equation}
The smallness of the singlet $vev$ $z$ is ensured by
the shinning mechanism, but it is constrained from
phenomenological considerations. 

The singly and doubly charged Higgs scalars in the
model could be in the range of TeV and their 
phenomenology has been extensively studied in the 
literature. The physical neutral Higgs scalars $H$ 
is the usual Higgs doublet of the standard model
and $\zeta$ is the physical neutral triplet Higgs
scalar with mass of about $m_\xi$, which may be
determined by the direct search at LHC or ILC. 

We shall now discuss the massless Nambu-Goldstone
Bosons, which may be represented by the normalized
fields $v \theta$, $u \eta$, and $z \varphi$, and
the physical mass eigenstates may be expressed as
\begin{equation}
\left(
\begin{array}{c}
J^0 \\ G^0 \\ \Omega^0
\end{array}
\right) = U 
\left(
\begin{array}{c}
z\varphi \\ u\eta \\ v\theta
\end{array}
\right)\,.
\label{mvsg}
\end{equation}
The unitary matrix $U$ diagonalizes the mass-squared matrix
of the pseudoscalar bosons. Appropriate normalization
would reveal that the Goldstone mode
\begin{eqnarray}
G^0=\frac{v^2 \theta + 2 u^2 \eta}{\sqrt{v^2 + 4 u^2}}\,,
\end{eqnarray}
becomes the longitudinal mode of the $Z$-boson and
\begin{eqnarray}
\Omega^0=\frac{\varphi - \eta + 2 \theta}{\sqrt {z^{-2} + u^{-2} + 4 v^{-2}}}
\,,
\end{eqnarray}
is the physical heavy pseudoscalar Boson, which is mostly
triplet.
\begin{equation}
J^0 = {(v^2 + 4 u^2) z^2 \varphi + v^2 u^2 \eta - 2 u^2 v^2 \theta \over 
	\sqrt {z^2(v^2 + 4 u^2)^2 + u^2 v^4 + 4 v^2 u^4}}.
\label{j0}
\end{equation}
is the massless Majoron, which now becomes a pNGB in
the presence of the explicit lepton number violating term
$V_{\rm exp}$ and have a tiny mass $m_s \sim 10^{-3}$~eV.
The decays of the physical
field $\zeta$ into two Majorons gives the most severe
constraint on $z$, and we get the condition
$u < z < v$ with $z$ to be greater than MeV. This is
consistent with the tiny mass $m_s \sim 10^{-3}$~eV
of the pNGB. 
 
We shall now discuss how this pNGB can give rise to the
attractive force between the neutrinos and form condensates,
which in turn, can explain the observed dark energy. 
From the theory of superconductivity it is well-known that an 
overall attractive interaction is needed to form a condensate 
of the particles~\cite{book:Fetter}. Moreover, in the SM there 
is attractive interaction is possible among the neutrinos. 
For example the $Z$-boson exchange among two left-handed 
neutrinos give rise to a repulsive interaction among them. 
However, new interactions with characteristic scale $\sim$ eV 
can result in attractive interaction among 
neutrinos~\cite{Caldi:1999db}. The condensate can be formed 
through a spin zero pairing of the left and right handed 
neutrino mediated by a scalar with tiny mass. In our case
the pNGB does this job. One can in principle write the 
effective four-fermi interaction among the neutrinos and from 
there use the mean-field approximation and the standard 
Nambu-Gor'kov formalism~\cite{Nambu:1960tm, Gorkov:1958} to 
derive the gap equation and the relevant energy gap, 
$\Delta$ thereof. We refer to~\cite{Kapusta:2004gi, 
Bhatt:2009wb, Dey:2017wwt} for a detailed discussion. The 
energy gap in this case is
$\Delta = e^{-a}\sqrt{2\Lambda/M_{\nu}}(3\pi^{2}n_{\nu})^{1/3}$,
where $\Lambda$ is the cut-off scale which can be taken as the 
neutrino decoupling $\sim$ MeV~\cite{Bhatt:2009wb}, $n_{\nu}$ is 
the typical neutrino number density ($\sim$ 110/c.c.), $a$ is a 
dimensionless parameter which is given by 
$4\pi^{2}/(\mathcal{C} M_{\nu}(3\pi^{2}n_{\nu})^{1/3})$
with $\mathcal{C}$ being the effective coupling strength. 
In this context there are two more important quantities, namely, 
the critical temperature $T_{c} \simeq 0.57\Delta$ and Pippard
coherence length $\tilde{\xi} = e^{a}/(\pi \sqrt{2\Lambda M_{\nu}})$. 
Note that when the interaction between the neutrinos are 
sufficiently strong $\tilde{\xi}$ is comparable to the inter-particle 
spacing.  
To have a numerical estimation, if we take as a benchmark value 
the mass of the neutrino $M_{\nu} \sim 0.01$ eV, we can have a 
coherence length $\tilde{\xi} \sim 0.1$ cm, effective coupling 
strength $\mathcal{C}\sim 10^{6}$ eV$^{-2}$, energy gap 
$\Delta \sim 2 \times 10^{-6}$ eV and a critical temperature 
$T_{c} \sim 0.01$ K. Also to satisfy these values we have to 
have $f_{ij} \lambda_{0} \sim 10^{-8}$ which follows from 
Eq.~(\ref{eq:numass}). 
The existence of a finite, non-zero, gap may provide the evidence for a condensate which can be taken as the dark energy candidate~\cite{Bhatt:2009wb}. Also, for our choice of parameters the Cooper pairs are formed below the neutrino mass scale which makes the dark energy density comparable to the Majorana mass of the neutrinos which is $\mathcal{O}(10^{-3})$ eV.
In summary we considered a novel scenario of incorporating neutrino masses in models of TeV scale large extra dimensions. In addition we can address the issue of dark energy by means of neutrino condensates without resorting to any dynamical field like the quintessence or accelerons. Lastly, we leave a rigorous parameter space analysis for future study.   

%%%%%%%%%%%%%     Acknowledgements    %%%%%%%%%%%%%%%%%%%
\noindent
{\bf Acknowledgements\,:} \\
We acknowledge Tirtha Sankar Ray for discussions at the initial stages of this work. UKD acknowledges the support from Department of Science and Technology (DST), Government of India under the Grant Reference No. SRG/2020/000283. US thanks DAE for the Raja Ramanna Fellowship and an associated research grant.

%%%%%%%%%%%%%%%%%   References %%%%%%%%%%%%%%%%%%%%%%%%%%%%%%%%%%%%

\bibliographystyle{JHEP}
\bibliography{ref.bib}
\end{document}